\documentclass[aps, prd, amsmath, floats,floatfix, twocolumn, superscriptaddress,nofootinbib, showpacs]{revtex4}

\usepackage{natbib}
\usepackage{graphicx}
\usepackage{dcolumn}
\usepackage{bm}
\usepackage{amssymb}
\usepackage{amsmath}
\usepackage{verbatim}
\usepackage{mathrsfs}
\usepackage{amsfonts}
\usepackage{latexsym}
\usepackage{epsfig}
\usepackage{color}
\usepackage{graphicx,subfigure}
\usepackage{units}
\usepackage{array}

\begin{document}

\preprint{APS/123-QED}

\title{Dwarf spheroidal galaxies and Bose-Einstein condensate dark matter}

\author{Alberto Diez-Tejedor}
\affiliation{Santa Cruz Institute for Particle Physics and Department of Physics, University of California,
Santa Cruz, CA, 95064, USA}

\author{Alma X. Gonzalez-Morales}
\affiliation{Santa Cruz Institute for Particle Physics and Department of Physics, University of California,
Santa Cruz, CA, 95064, USA}

\author{Stefano Profumo}
\affiliation{Santa Cruz Institute for Particle Physics and Department of Physics, University of California,
Santa Cruz, CA, 95064, USA}

\begin{abstract}

We constrain the parameters of a self-interacting massive dark matter scalar 
particle in a condensate using the kinematics 
of the eight brightest dwarf spheroidal satellites of the Milky Way.
For the case of a repulsive self-interaction the condensate
develops a mass density profile with a characteristic scale radius 
that is closely related to the fundamental parameters of the theory.
We find that the velocity dispersion of dwarf spheroidal galaxies
suggests a scale radius of the order of $1\,\textrm{kpc}$, in tension with previous
results found using the rotational curve of low-surface-brightness and dwarf galaxies.
The new value is however favored marginally by the constraints coming from the number of relativistic species at Big-Bang nucleosynthesis.
We discuss the implications of our findings for the particle dark matter model and argue that while a single classical coherent state can correctly describe the 
dark matter in dwarf spheroidal galaxies, it cannot play, in general, a relevant role for the description 
of dark matter in bigger objects.

\end{abstract}


\maketitle

\section{Introduction}

The nature of dark matter (DM) remains an open question.  At the fundamental level, DM is expected to be described in terms of a quantum field theory. 
At the effective level, however, a description in terms of classical particles is usually considered,
see e.g. the large literature on N-body simulations~\cite{Springel:2005nw}. Most current efforts are focused on detecting a 
weakly interacting massive particle (WIMP), both by direct~\cite{direct} and indirect~\cite{indirect} searches. 
In the case of WIMPs its present-day abundance is fixed at the time when DM decoupled from the thermal plasma. 
If the interaction of DM lies at the weak scale, with a mass of the particle in the range of $100\,\rm GeV$ (as expected from the supersymmetric extensions to 
the standard model), the energy density of these particles coincides ``miraculously'' with the observed one~\cite{WIMP}.
However, alternatives exist and deserve careful scrutiny, either to constrain the associated parameter space, and thus phenomenology,
or to dismiss them as viable candidates. 

One such proposal considers that the abundance of DM
is fixed by an asymmetry between the number densities of particles and antiparticles~\cite{asymmetricDM}, 
similarly to the baryons and leptons in the universe. If the particle interactions in the early universe 
are strong enough to guarantee thermal equilibrium, and DM is further composed of a spin-0 quantum field, the zero mode could have developed a Bose-Einstein
condensate where a description in terms of a classical field  
would be warranted. Classical coherent states can also emerge nonthermally, no  asymmetry required, by means of the vacuum misalignment mechanism~\cite{misalignment}.
Similar ideas have been considered previously in the literature under many different names, such as scalar field~\cite{Suarez:2013iw},
BEC~\cite{Boehmer:2007um}, Q-ball~\cite{Qball}, fuzzy~\cite{Hu:2000ke}, 
boson~\cite{Sin-Bose}, or even fluid~\cite{Peebles:2000yy}, DM; see also 
Refs.~\cite{BECgeneral, Sikivie:2009qn, Arbey:2002, Arbey:2003sj, Lee:1995af, Harko.2011, Robles.2012, Dwornik:2013fra, Hua, Lora, Lora2} for details.

A natural realization of this scenario can be provided by the axion~\cite{Sikivie:2009qn}. Originally introduced to solve the charge-parity violation 
problem in QCD~\cite{Peccei:1977hh}, the axion was soon recognized as a promising candidate for DM. In this case the size of the condensate 
is so small~\cite{Barranco:2010ib} that, most probably, DM halos made of axion-balls
could not be distinguished from the ones simulated with N-body codes by means of galactic dynamics and/or lensing observations~\cite{GonzalezMorales:2012ab}.
Another possibility is that with an appropriate choice of the parameters in the model (see the next two paragraphs for details), 
it could be possible to develop single structures with the size of a galaxy~\cite{Boehmer:2007um, Sin-Bose, Lee:1995af, Arbey:2002, Arbey:2003sj, Harko.2011, Robles.2012, Dwornik:2013fra, Hua, Lora}.

For practical purposes we will restrict our attention to the case of a massive, self-interacting, complex scalar 
field with an internal $U(1)$ global symmetry satisfying the Klein-Gordon equation
\begin{equation}\label{eq:KG}
\Box \phi-(mc/\hbar)^2\phi-2\lambda|\phi|^2\phi=0\,.
\end{equation}
Here the box denotes the d'Alembertian operator in four dimensions, with $m$ the mass of the scalar particle
and $\lambda$ a dimensionless self-interaction term. As long as the interaction between bosons is repulsive, $\lambda >0$,
a universal mass density profile for the static, spherically symmetric, regular, asymptotically flat, self-gravitating equilibrium scalar field configurations
emerges in the weak field, Thomas-Fermi regime~\cite{Colpi:1986ye, Lee:1995af, Arbey:2003sj, Boehmer:2007um, Harko.2011} 
of the Einstein-Klein-Gordon system with the following analytic form:
\begin{equation}
  \label{density}
  \rho(r)= \left\lbrace 
    \begin{array}{cl}
      \rho_c\, \displaystyle{\frac{\sin(\pi r/r_{\textrm{max}})}{(\pi
          r/r_{\textrm{max}})}} \quad & \textrm{for} \quad
      r<r_{\textrm{max}} \; \\
      0  \quad & \textrm{for} \quad r\ge r_{\textrm{max}} \; 
    \end{array}
  \right. \; .
\end{equation}
In the effective description above there are two free parameters: first, the size of the gravitating objects, 
\begin{equation}\label{eq:rmax}
 r_{\textrm{max}}\equiv\sqrt{\frac{\pi^2 \Lambda}{2}}\left(\frac{\hbar}{m c}\right)= 48.93 \left(\frac{\lambda^{1/4}}{m[\textrm{eV}/c^2]}\right)^2 \,\textrm{kpc}\,,
\end{equation}
a parameter that, as manifest from the equation, depends directly on the bare constants 
of the theory in the combination $m/\lambda^{1/4}$; second, the value of the mass density at the center of the configuration, 
$\rho_c\equiv\pi m Q/(4r_{\textrm{max}}^3)$, a quantity that in principle can vary from galaxy to galaxy.
Here $Q$ is the total charge in the system, that in this case coincides with the total number of particles, and for convenience we have
defined the dimensionless constant $\Lambda\equiv \lambda m_{\textrm{Planck}}^2/4\pi m^2$.

The mass density profile in Eq.~(\ref{density}) describes only the diluted configurations of a scalar field in a regime of weak gravity; in terms of 
particle numbers that translates into (see e.g. Eq.~(25) in Ref.~\cite{Arbey:2002})
\begin{equation}\label{eq:inequalities}
\Lambda^{-1/2}\ll \left(\frac{m}{m_{\textrm{Planck}}}\right)^2Q\ll \Lambda^{1/2}\,.
\end{equation}
The inequalities in Eq.~(\ref{eq:inequalities}) demand $\Lambda\gg1$; that is guaranteed if the 
combination $m/\lambda^{1/4}$ for the mass and self-interaction terms of the scalar boson 
is well below the Planck scale.
It is precisely the very large value expected for the constant $\Lambda$ what makes possible
to blow up the Compton wavelength of the scalar particle, $\hbar /m c$, up to galactic scales, see Eq.~(\ref{eq:rmax}) above.
Only configurations with masses $M=mQ$ in the range from $M\gg \lambda^{-1/2}m_{\textrm{Planck}}$ up to $M\ll \lambda^{1/2}m^3_{\textrm{Planck}}/m^2$ 
can be described by the expression in Eq.~(\ref{density}). 
Then, in order to have an halo model for objects of at least $M\sim 10^{8}M_{\odot}$ ($M\sim 10^{12}M_{\odot}$) we
need a scalar DM particle with 
$m/\lambda^{1/4}< 70\,\textrm{keV}/c^2$ ($m/\lambda^{1/4}< 0.7\,\textrm{keV}/c^2$).

\begin{figure*}[]
 \includegraphics[scale=0.7]{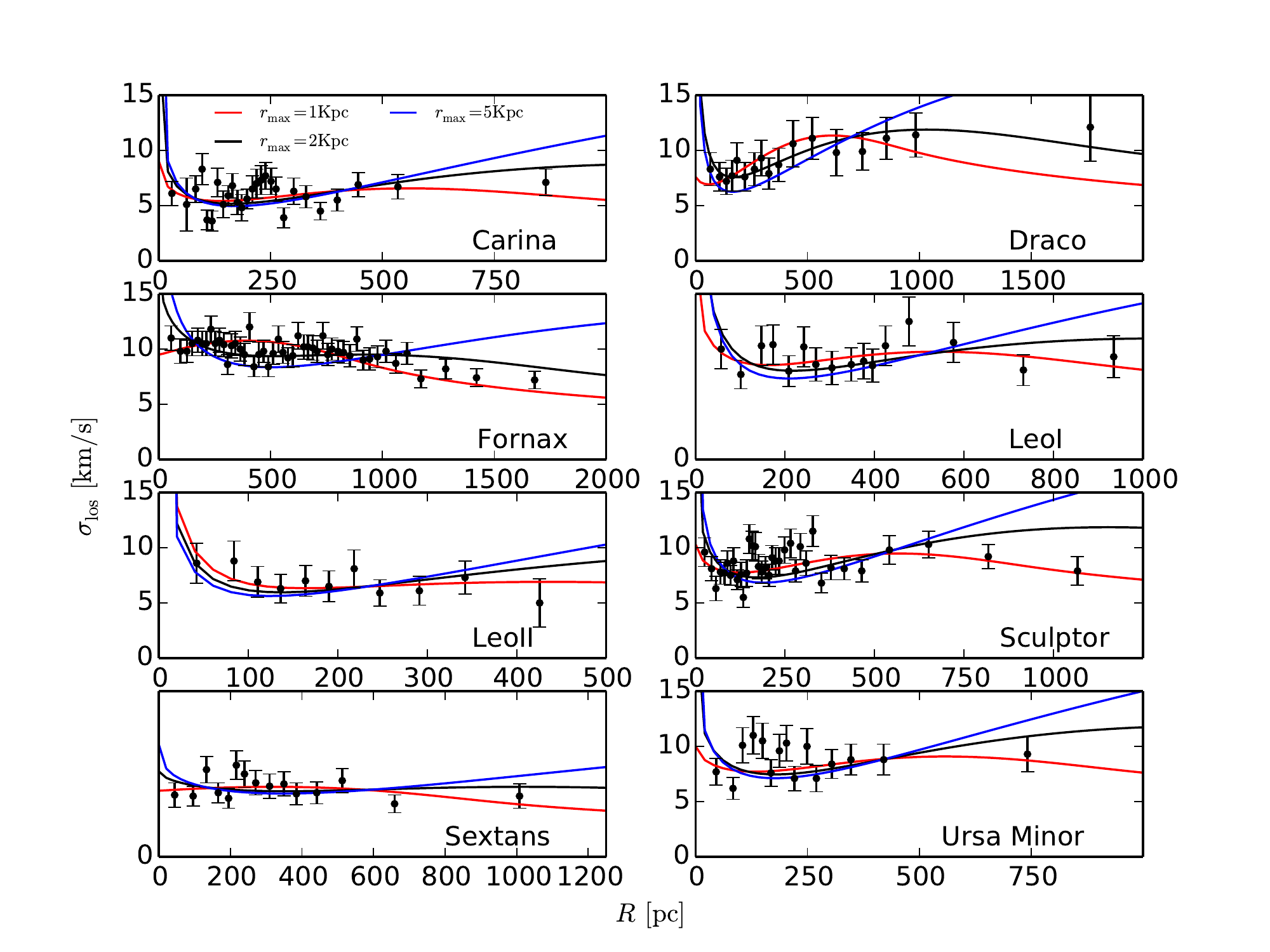} 
 \caption{Empirical, projected velocity dispersion profiles for the classical eight dSph satellites of the Milky Way as reported in 
 Refs.~\cite{Walker:2009zp, Walker:2007ju, Mateo2008}. Solid lines denote the best fits for 
 the halo model in Eq.~(\ref{density}) when $r_{\textrm{max}}=1\,\textrm{kpc}$ (red), $r_{\textrm{max}}=2\,\textrm{kpc}$ (black), and 
 $r_{\textrm{max}}=6\,\textrm{kpc}$ (blue). }
 \label{fig:fig1}
\end{figure*}

The density profile in Eq.~(\ref{density}) was derived under the assumption that all the DM particles are in a condensate, 
while in a more realistic situation probably only a fraction of them would be represented by the coherent classical state. 
(That seems indeed necessary in order to explain the flattened rotation curves in large spirals, where observations suggest 
$\rho\sim 1/r^2$ at large radii.)
Unfortunately, there is not yet a satisfactory description that includes this effect 
(see Ref.~\cite{Bilic:2000ef} for a proposal in this direction). Nevertheless, 
this halo model can still be deemed appropriate to test the self-interacting scalar field DM scenario
if we carefully choose observations that are sensitive only to the mass contained up to a radius smaller or comparable to 
$r_{\rm max}$, where the condensate is expected to dominate the distribution of DM.
One should then look at the profile in Eq.~(\ref{density}) not necessarily as a DM halo model for the whole galaxy, but  
for the core of the self-gravitating object only.

The dwarf spheroidal (dSph) satellites of the Milky Way are probably the most promising objects to test DM models as far as structure formation is concerned. These old, 
pressure-supported systems are the smallest and least luminous known galaxies, and there is strong evidence that they are DM-dominated at all radii, with mass-to-light 
ratios as large as \cite{Mateo:1998wg} 
\begin{equation}
M/L_V\sim 10^{1-2}[M/L_V]_{\odot}\,.
\end{equation}
The dynamics of these objects, 
for instance, could allow us to determine whether DM halos are cored or cuspy: since the 
concentration of baryons in these galaxies is so low, effects such as the adiabatic contraction and/or supernova
feedback cannot alter significantly the shape of the original halo. 
Current data do not yet conclusively discriminate between cuspy and cored profiles~\cite{Walker:2009zp, Salucci:2011ee, jardel, Valenzuela:2005dh}.

In this paper we use the kinematics of the eight classical dSph satellites of the Milky Way
to determine whether a self-interacting scalar particle in a condensate is able to reproduce the galaxies' internal dynamics 
and, if so, under what conditions on the theory input parameters. In this respect, our study extends 
previous analyses carried out for the generalized Hernquist~\cite{Walker:2009zp}
and Burkert~\cite{Salucci:2011ee} profiles to the DM halo model in Eq.~(\ref{density}).
It is important to note, however, that the purpose of this paper is not to compare the profile in Eq.~(\ref{density}) 
with other halo models in the literature, 
but, rather, to use dSph dynamics to test 
the self-consistency of the scalar field dark matter scenario. 

We find that the eight classical dSphs indicate a scale radius of the order
\begin{equation}
r_{\textrm{max}}\sim 1\, \textrm{kpc}\,,\quad {\rm i.e.}\quad m/\lambda^{1/4}\sim 7\,\textrm{eV}/c^2\,,
\end{equation}
a value in tension with previous results found using the rotation curves of low-surface-brightness 
(LSB) and dwarf galaxies~\cite{Boehmer:2007um, Arbey:2003sj, Harko.2011, Robles.2012}; see also Refs.~\cite{Dwornik:2013fra, Hua} for bigger galaxies. 
Our findings strongly disfavor a {\it self-interacting condensate DM halo model} or,
if one hypothesizes that the condensate describes only the core of galaxies, 
they indicate that the relevance of the coherent state to describe DM in larger galaxies is, at best, 
negligible.

\section{The Jeans equation}

Dwarf spheroidal galaxies are simple, old systems composed of a DM halo and of a stellar population.  
Rotation in these galaxies is negligible, and the stellar component 
is supported against gravity by its random motion. Therefore 
the observation that can be used to test DM models is not rotation curves but, rather, the line-of-sight velocity dispersions. 

Walker {\it et al}~\cite{Walker:2009zp, Walker:2007ju, Mateo2008} reported updated empirical velocity dispersion profiles for the  eight ``classical'' dSphs
of the Milky Way: Carina, Draco, Fornax, Leo I, LeoII, Sculptor, Sextans, and Ursa Minor; see Figure~\ref{fig:fig1} for details.
Following standard parametric analysis~\cite{Walker:2009zp, Salucci:2011ee} (see Ref.~\cite{jardel} for a different approach), 
we consider that the stellar component in each individual galaxy is in dynamical equilibrium and that it traces the underlying DM distribution. 
Assuming, further, spherical symmetry, Jeans's equation relates the mass profile of the DM halo, 
\begin{equation}
M(r)=\frac{M_{\rm max}}{\pi} \left[\sin\left(\frac{\pi r}{r_{\rm max}}\right)-\frac{\pi r}{r_{\rm max}}\cos\left(\frac{\pi r}{r_{\rm max}}\right)\right]\,,
\end{equation}
where 
\begin{equation}
M_{\rm max}=M(r_{\rm max})=(4/\pi)\rho_c r_{\rm max}^3\,,
\end{equation}
to the first moment of the stellar distribution function, 
\begin{equation}
\frac{1}{\nu}\frac{d}{dr}\left(\nu\langle v^2_r\rangle\right)+2 \frac{\beta\langle v^2_r\rangle}{r} = -\frac{G M}{r^2}\,.
\end{equation}
Above, $\nu(r)$, $\langle v^2_r (r)\rangle$, and $\beta(r)=1-\langle v^2_{\theta}\rangle/\langle v^2_r\rangle$ are the three-dimensional 
density, radial velocity dispersion, and orbital anisotropy, respectively, of the stellar component.
The parameter $\beta$ quantifies the degree of radial stellar anisotropy:  
if all orbits are circular $\langle v^2_r\rangle = 0$, and then $\beta = \infty$; if the orbits are isotropic 
$\langle v^2_r\rangle=\langle v^2_{\theta}\rangle$, and $\beta= 0$; finally, if all orbits are perfectly radial, 
$\langle v^2_{\theta}\rangle = 0$, then $\beta = 1$.
There is no preference {\it a priori} for either radially, $\beta >0$, or tangentially, $\beta<0$, biased systems; however, 
configurations with $\beta \sim 1$ are disfavored due to the very particular initial conditions they seem to require.

In the simplest scenario with constant orbital anisotropy, $\beta(r)=\textrm{const}$, the (observed) projection of the velocity 
dispersion along the line-of-sight, $\sigma_{\textrm{los}}^2 (R)$, relates the mass profile, $M(r)$, to the (observed) stellar density, $I(R)$, 
through~\cite{Binney}
\begin{equation}
  \sigma_{\textrm{los}}^2=\frac{2G}{I(R)} \int_R^{\infty}{dr' \nu(r') M(r')(r')^{2\beta-2} F(\beta,R,r')} \,.
  \label{eq:sigmalos}
\end{equation}
Here
\begin{equation}
F(\beta,R,r')\equiv \int_R^{r'}{dr \left(1-\beta\frac{R^2}{r^2}\right) \frac{r^{-2\beta +1}}{\sqrt{r^2-R^2}}}\,,
\end{equation}
and $R$ is the projected radius. We adopt a Plummer profile for the stellar density,
\begin{equation}\label{eq.Plummer}
I(R)=\frac{L}{\pi r_{\rm half}^2}\frac{1}{[1+(R/r)^2)]^2}\,,
\end{equation}
where $L$ is the total luminosity of the object and $r_{\rm half}$ (the only single shape parameter) 
the half-light radius. The values of these two quantities for each of the eight classical dSphs are listed in Table~I of Ref.~\cite{Walker:2009zp}.
Under the assumption of spherical symmetry the corresponding three-dimensional stellar density associated with the 
Plummer profile takes the form 
\begin{equation}
\nu(r)=\frac{3 L}{4\pi r_{\rm half}^3}\frac{1}{[1+(r/r_{\rm half})^2]^{5/2}}\,.
\end{equation}
We have corroborated that our findings in this paper are not very sensitive to the profile of the stellar component, and
similar results are also obtained using a Sersic~\cite{Sersic} or a King profile~\cite{King}.

\begin{figure*}
 \includegraphics[width=0.32\textwidth]{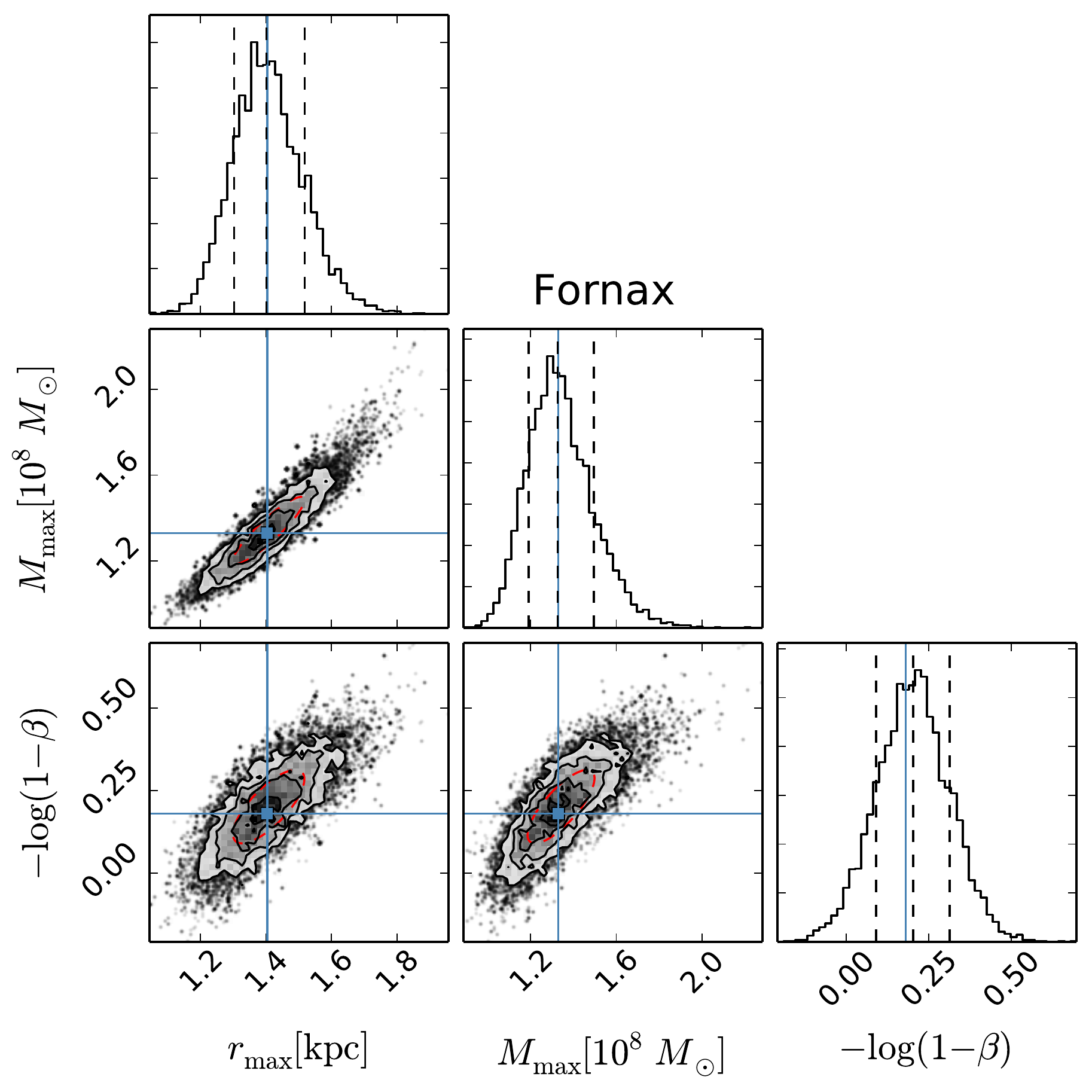}  
 \includegraphics[width=0.32\textwidth]{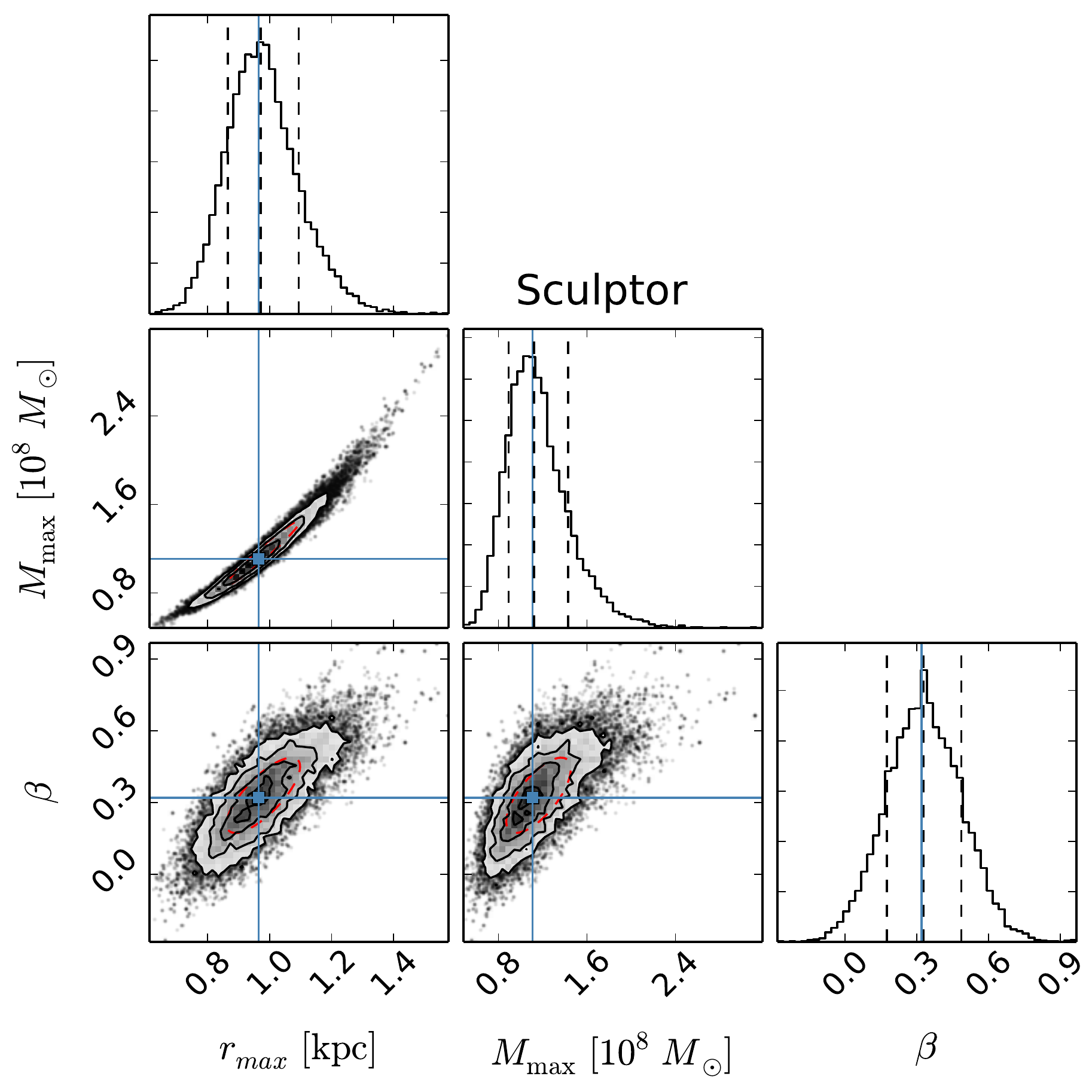} 
 \includegraphics[width=0.32\textwidth]{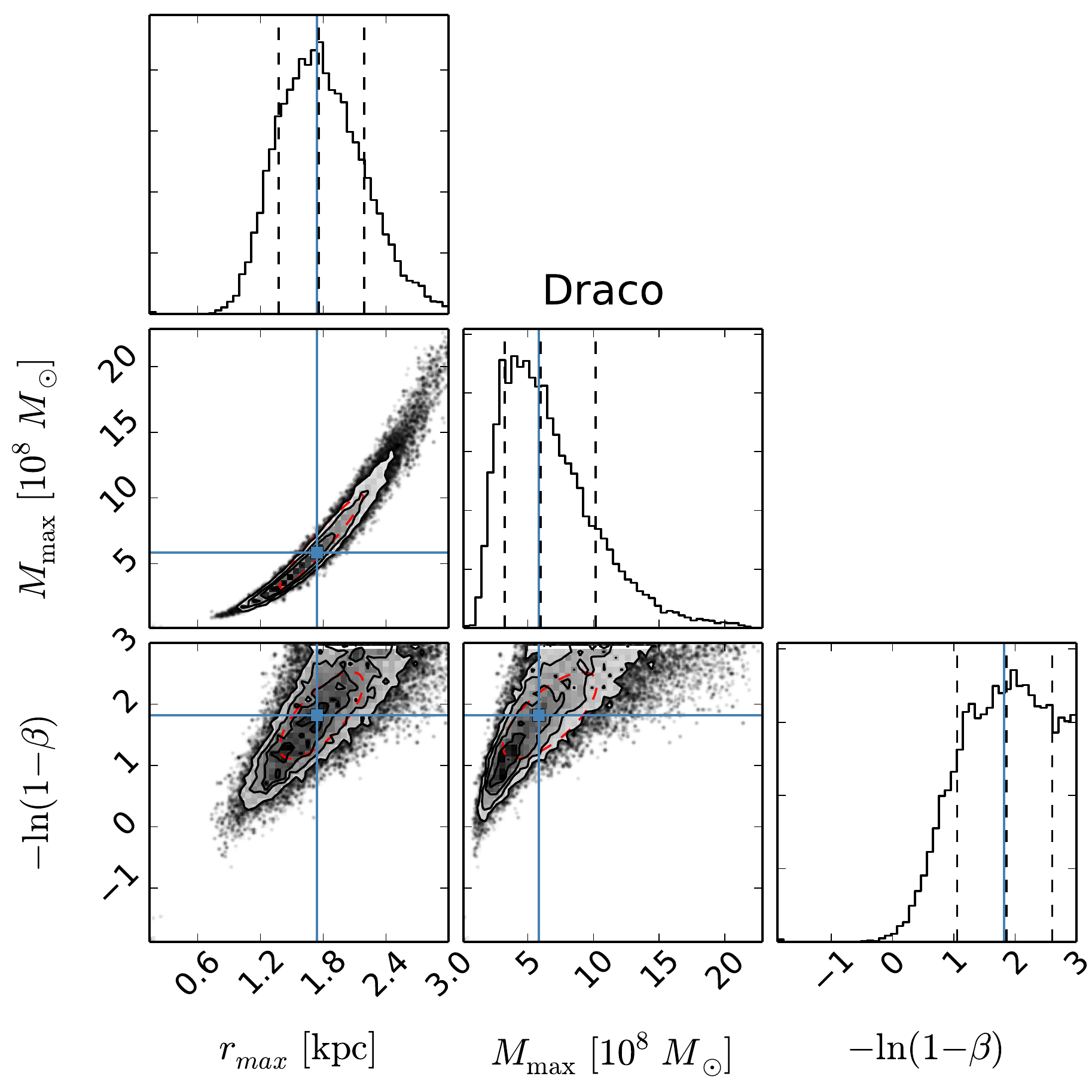}
 \caption{Two-dimensional posterior distributions of Fornax, Sculptor, and Draco using the BEC halo model in Eq.~(\ref{density}).  
 The histograms correspond to the marginalized posterior distributions of each parameter. The dashed lines and red contours represent the $1\sigma$ 
 confidence interval. Solid lines indicates the maximum likelihood point.}
 \label{fig:MCMC1}
\end{figure*}

\subsection{Maximum Likelihood and Monte Carlo analysis}

In order to fit the observations we have three free parameters per galaxy: two associated with the halo model, the scale radius $r_{\rm max}$ and the total mass 
$M_{\rm max}$, and one associated with the stellar component, the orbital anisotropy $\beta$. Since the scale radius is a constant in the theory one could perform a 
combined analysis for the eight galaxies keeping this quantity fixed. For the purpose of this paper, however, this procedure is not warranted; instead we 
estimate $r_{\rm max}$ for each galaxy, and we then compare the values obtained for the  
different galaxies. We will also contrast our results against previous constraints arising from the study of the rotational curves of LSB and dwarf galaxies.
As we show below, this analysis is sufficient to uncover strong tension between model and  observations at different scales.

In order to proceed we perform a Maximum Likelihood$-$Markov chain Monte Carlo analysis (we use the EMCEE code, described in Ref.~\cite{Foreman-Mackey}) 
to explore the parameter space and estimate the values of $r_{\rm max}$, $M_{\rm max}$ and $\beta$ for each individual galaxy, 
together with their corresponding uncertainties. For each galaxy we define the likelihood function
\begin{equation}
 \mathcal{L}=\prod_{i=1}^N{\frac{\exp \left[-\frac{1}{2}\frac{\left(\sigma_{\rm los}^{\rm obs}(R_i)-\sigma_{\rm los}(R_i)\right)^2}{{\rm Var}[\sigma_{\rm los}^{\rm obs}(R_i)]}
 \right]}{ \sqrt{ 2 \pi \, {\rm
Var}[\sigma_{\rm los}^{\rm obs}(R_i)]}}}\,.
\end{equation}
Here $\sigma_{\rm los}^{\rm obs}(R_i)$ is the observed line-of-sight velocity dispersion at projected radius $R_i$, 
$\sigma_{\rm los}(R_i)$  is given in Eq.~\eqref{eq:sigmalos}, 
${\rm Var}[\sigma_{\rm los}^{\rm obs}(R_i)]$ is the square of the error associated with the observed value of the velocity dispersion at $R_i$, 
and $i$ is a label for the data bins that runs from $1$ to the total number of bins $N$. To account for the uncertainties on $r_{\rm half}$  
we marginalize over this parameter by sampling it, at each step of the Monte Carlo, from a normal distribution with a 
standard deviation equal to its actual uncertainty.

For the three free parameters we adopt uniform log-priors in the following ranges:
\begin{subequations}
\begin{eqnarray}
-2.5<&\ln \left(r_{\rm max}\,[\rm kpc]\right)&<2.5 \,, \\
-7<&\ln \left(M_{\rm max} \,[10^9 \, M_{\odot}]\right)&<7 \,, \\
-3<&-\ln \left(1-\beta\right)&<3  \,.
\end{eqnarray}
\end{subequations}
For each galaxy we run 50 chains simultaneously, starting at random values within the prior range, and allow each chain to run for 1,000  steps, 
from which we eliminate the first 100 steps that correspond to a ``burn-in'' period.

\section{Results}

Our results are shown in Figure~\ref{fig:MCMC1} where, for three of the galaxies  
with more data points, Fornax, Sculptor and Draco, we plot the one- and two-dimensional posterior distributions of the  
parameters $r_{\rm max}$, $M_{\rm max}$, and $\beta$. 
As we can note the posterior distributions are almost symmetric with respect to the maximum likelihood point (solid lines). The dashed lines and
red ellipses indicate the 1$\sigma$ (68.2\% C.L.) confidence interval of the different parameters.
Some degeneracy between the scale radius and the total mass, and the anisotropy, is evident; 
however, in all cases the chains converge to a small region of the parameter space. 

The values of $r_{\rm max}$, $M_{\rm max}$ and $\beta$ for all the galaxies in the sample,  
together with their corresponding uncertainties, are listed in Table~\ref{tab:results}.  
We have corroborated that similar results are also obtained when using a Sersic (King) stellar
distribution. In particular, for $r_{\rm max}$ we obtain a difference of $\sim 0.5\,{\rm kpc}$ ($\sim 0.2 \rm kpc$) in the central value, 
but the error remains of the same magnitude with respect to that in the Plummer case.

\begin{table}
\begin{tabular}{lcccc}
\\\hline\hline
$\;$Object$\quad$  & $\;\; r_{\textrm{max}}[\textrm{kpc}] \;\;$ & $\;\;M_{\textrm{max}}[10^8 M_{\odot}]\;\;$ & $\;\;-\ln ( 1-\beta)\;\;$ \\ \hline
$\;$Fornax   & $1.4\substack{+0.1 \\ -0.1}$ & $1.1 \substack{+0.9 \\ -0.9} $ &$ 0.2 \substack{+0.1 \\ -0.1}$ \\
$\;$Sculptor  & $1.0\substack{+0.1 \\ -0.1}$ &$1.1 \substack{+0.3 \\ -0.2}$ & $0.3\substack{+0.2 \\ -0.2}$ \\

$\;$Carina    & $1.1 \substack{+0.3 \\ -0.3}$ & $0.8 \substack{+0.6 \\ -0.3}$& $0.6 \substack{+0.3 \\ -0.3}$ \\
$\;$Draco   & $1.7\substack{+0.4 \\ -0.3}$ & $5.9\substack{+4.1 \\ -2.7} $ & $1.8 \substack{+0.7 \\ -0.8} $ \\
$\;$Leo I   & $ 1.0 \substack{+0.4 \\ -0.2}$ & $1.7 \substack{+1.4 \\ -0.7}$& $0.9 \substack{+0.7 \\ -0.5}$ \\
$\;$Leo II   & $0.6 \substack{+0.3 \\ -0.2}$ & $0.5 \substack{+0.7 \\ -0.3} $ & $1.6 \substack{+0.9 \\ -0.9} $ \\
$\;$Sextans   &  $0.7 \substack{+0.4 \\ -0.3}$ & $0.2 \substack{+0.2 \\ -0.1}$ & $ -0.4 \substack{+0.4 \\ -0.7}$ \\
$\;$Ursa Minor   &  $0.9 \substack{+0.4 \\ -0.3}$& $0.9 \substack{+0.9 \\ -0.4}$ & $ 0.1 \substack{+0.3 \\ -0.3}$ \\
\hline
\end{tabular}
\caption{Estimate of the parameters $r_{\rm max}$, $M_{\textrm{max}}$, and $\beta$ for the classical dSphs in the Milky Way.} 
\label{tab:results}
\end{table}

We conclude that the preferred value of the scale radius  inferred from the dynamics of the eight dSphs
lies around $r_{\rm max}\sim 1 \,\textrm{kpc}$, i.e. $m/\lambda^{1/4}\sim 7\,\textrm{eV}/c^2$; 
this value is indeed contained within the 3$\sigma$ (99.7\% C.L.) confidence interval of each galaxy.  Moreover, we can exclude at
more than 5$\sigma$ (99.9\% C.L.) values of $r_{\rm max}\gtrsim 5 \,{\rm kpc}$. As we will discuss next in Section~\ref{sec:discussion}, 
this implies a strong conflict with previous constraints on this parameter
of the theory. 

\begin{figure}[t!]
 \includegraphics[width=0.48\textwidth]{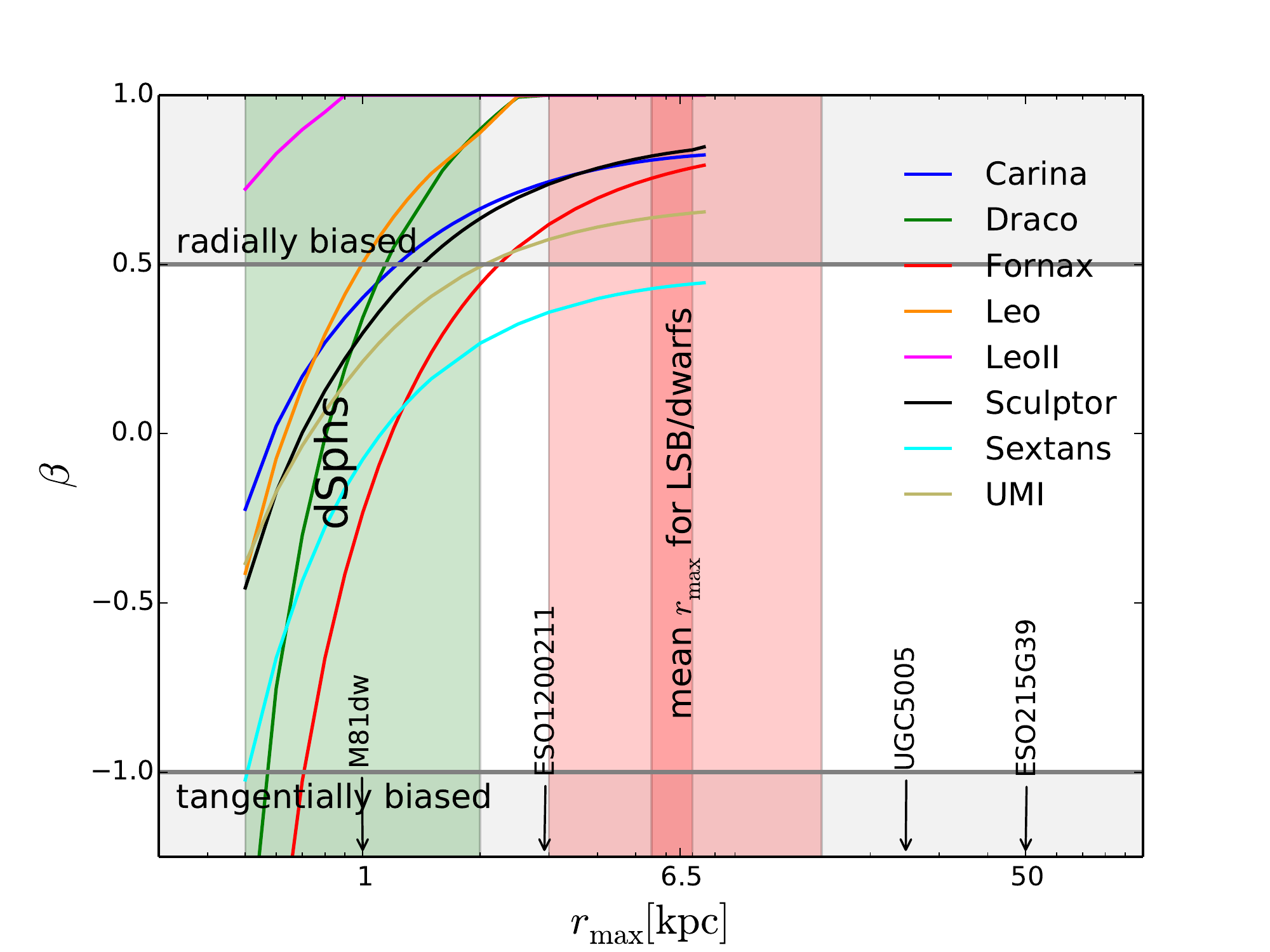} 
 \caption{Preferred orbital anisotropy for the best fits as a function of the scale radius. The lines at $\beta=0.5$
 and $\beta=-1$ correspond to $\langle v^2_r\rangle = 2\langle v^2_{\theta}\rangle $ and 
 $\langle v^2_{\theta}\rangle = 2\langle v^2_r\rangle$, respectively.}
 \label{fig:fig3}
\end{figure}

At this point we would like to stress that, besides the statistical evidence for small values of the parameter $r_{\rm max}$, there are also 
physical arguments that support this conclusion, which we can draw by looking at the  behavior of the best
fit parameters (minimum chi-square) of the anisotropy, $\beta$, and total mass, $M_{\rm max}$, 
for a fixed value $r_{\rm max}$ of the size of the condensate:
 
$(i)$ Density profiles with scale radii larger than $2\,\textrm{kpc}$ imply values of the anisotropy parameter 
$\beta\gtrsim 0.5$; see Figure~\ref{fig:fig3}. For a scalar field DM model there is no known connection between 
the anisotropy in the stellar distribution and the halo, so that dSphs could in principle be described as equilibrium 
systems even with such large values of the orbital anisotropy. 
(It is unclear to us whether large values of the stellar anisotropy would necessarily develop a radial instability for these halo models.) 
However, although these configurations cannot be excluded {\it a priori},  
they imply an unnatural preference for radial orbits.

\begin{figure}[!ht]
 \includegraphics[width=0.48\textwidth]{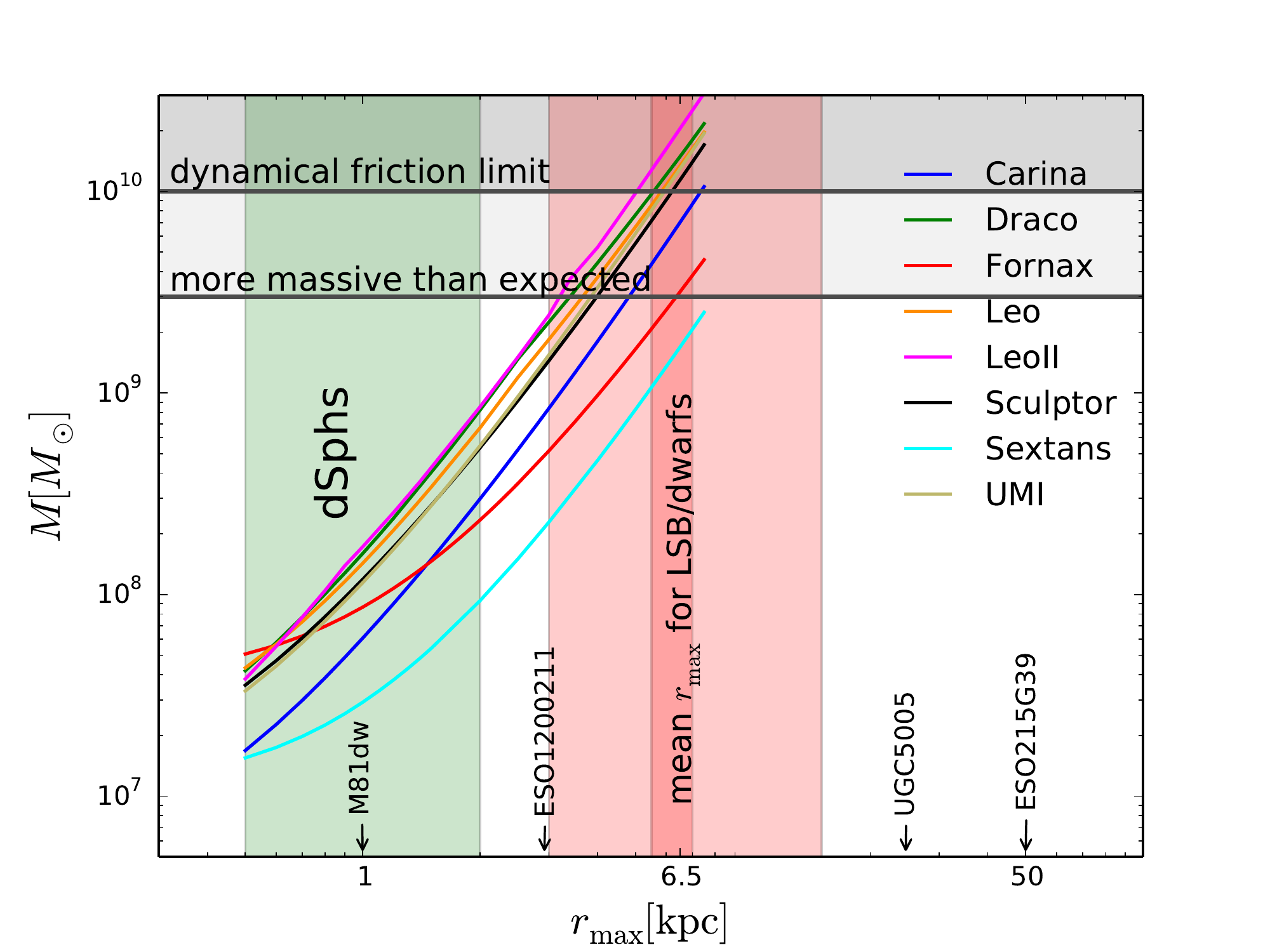} 
 \caption{Total mass for the best fits as a function of the scale radius. 
 The line at $M=3\times 10^9 {M_{\odot}}$ corresponds to the virial mass of Draco (the most massive object in the sample) obtained from a NFW profile  
 consistent with the observations in the velocity dispersions~\cite{Walker:2007ju, Jardel:2012am}. 
 The line at $M=1\times 10^{10}{ M_{\odot}}$ comes from an upper limit to the mass of this same galaxy
 as required from the dynamical friction decay time to be larger than one Hubble time~\cite{Binney, Gerhard}.
 }
 \label{fig:fig4}
\end{figure}

$(ii)$ As the value of the scale radius increases, the total mass required to fit the data grows drastically, reaching values as large as
$M_{\textrm{max}}\gtrsim 10^{10} M_{\odot}$ in some cases when $r_{\textrm{max}}\gtrsim 6\,\textrm{kpc}$; see Figure~\ref{fig:fig4}. This value is an
order of magnitude larger than what inferred by previous analysis~\cite{Walker:2009zp, Salucci:2011ee, Walker:2007ju, Jardel:2012am}. 
An upper limit to the mass of these objects 
stems from the requirement that the dynamical friction decay time not be larger than the age of the universe~\cite{Binney, Gerhard},
although there are no model independent limits on the total mass of these galaxies. 

Finally, it is also interesting to note that observations suggest a decline in the velocity dispersion profiles at large projected radii~\cite{Walker:2009zp, Lokas2008}, 
whereas the predicted profiles for large values of the scale radius grow at large radii.  
Even though for some galaxies the fit is not drastically worsen for large values of $r_{\textrm{max}}$,
if we inspect the overall radial dependence we can see that large scale radii fail in describing the outer regions 
for all galaxies, see the blue lines in  Figure \ref{fig:fig1}. 

From the above considerations the preference of a scale radius in the range $r_{\textrm{max}}\sim 0.5-2\,\textrm{kpc}$ 
(green band in Figures~\ref{fig:fig3} and~\ref{fig:fig4}) is solid.
A common value of $r_{\textrm{max}}$ larger than $5\,\textrm{kpc}$ is clearly disfavored by the dynamics of dSphs.

\section{Discussion and conclusions}\label{sec:discussion}

The viability of the halo model in Eq.~(\ref{density}) has been studied in several papers mainly employing rotational curves of galaxies from different surveys, 
out of which only the most DM-dominated objects have been selected~\cite{Boehmer:2007um, Arbey:2003sj, Harko.2011, Robles.2012}; 
see also Ref.~\cite{Lora, Lora2} for a different approach. 
These studies all point to a scale radius that varies from galaxy to galaxy and ranges from $3\,\textrm{kpc}$ up to $15\,\textrm{kpc}$ (light red band in 
Figures~\ref{fig:fig3} and~\ref{fig:fig4}),
with only isolated instances requiring values 
outside this range, e.g. M81dw, where $r_{\textrm{max}}\sim 1\,\textrm{kpc}$~\cite{Boehmer:2007um}, and UGC5005, where 
$r_{\textrm{max}}\sim 24.65\,\textrm{kpc}$~\cite{Harko.2011}.  
However, these papers also report mean values in the narrow range $r_{\textrm{max}}\sim 5.5-7\,\textrm{kpc}$~\cite{Arbey:2003sj, Harko.2011} 
(red band in Figures~\ref{fig:fig3} and~\ref{fig:fig4}), suggesting
the existence of a self-interacting scalar particle with $m/\lambda^{1/4}\sim 2.6-2.9\,\textrm{eV}/c^2$.
Such findings have led to the conclusion that the halo model in Eq.~(\ref{density}) can describe accurately the dynamics of DM-dominated galaxies. 
The case of Milky Way-like systems, or giant ellipticals, remains to be studied in detail mainly
because the dynamical interaction between the condensate and baryons is not well understood there (see however Ref.~\cite{Dwornik:2013fra},
where a set of three high-surface-brightness spirals have been recently considered, e.g. ESO215G39, where $r_{\textrm{max}}\sim 50\,\textrm{kpc}$,
and Ref.~\cite{Hua}, where values of the scale radius in the range
$r_{\textrm{max}}=5.6 - 98.2\,\textrm{kpc}$ are reported for a subsample of galaxies in the THINGS survey).
Note that, contrary to other proposals in the literature, the halo model in Eq.~(\ref{density}) is not expected to describe galaxy clusters.

The values reported in previous studies are strongly disfavored by our findings in the present analysis, where we show that the dynamics of the smallest 
and least luminous galaxies is clearly in conflict, along several lines, with such large scale radii.
One could argue that the profile in Eq.~(\ref{density}) is not appropriate to describe the galaxies in 
Refs.~\cite{Boehmer:2007um, Arbey:2003sj, Harko.2011, Robles.2012, Dwornik:2013fra, Hua} 
(where in some cases the luminous matter extends up to $10\,\textrm{kpc}$), 
and suggest that a more elaborated halo model 
where the condensate represents only the core of the galaxy 
would be necessary in order to understand the dynamics of these systems.
However, it is important to note that a condensate with a scale radius of the order of $1\,\textrm{kpc}$
does not provide the core expected for those galaxies used in previous analysis.

Interestingly, the new value of $r_{\textrm{max}}\sim 1\,\textrm{kpc}$ is favored by cosmological observations. A homogeneous and isotropic distribution of matter
satisfying the Eq.~(\ref{eq:KG}) has two different regimes depending on the actual value of the charge density, $q=Q/a^3$; see e.g. Ref.~\cite{Arbey:2002}. Here $a$
is the scale factor and $Q$ the number of particles per unit volume today, $a=1$. 
When the charge density is high, $q\gg m^3c^3/(\lambda \hbar^3)$,
the energy density and pressure of the scalar field dilute with the cosmological expansion like dark radiation, $\rho=3\lambda^{1/3} Q^{4/3} c\hbar/(4a^4)$
and $p=(1/3)\rho$, whereas at low densities, when $q\ll m^3c^3/(\lambda\hbar^3)$, like cold DM, $\rho= Qmc^2/a^3$ and $p=0$.
From the condition that the transition from dark radiation to DM, fixed at $q\sim m^3c^3/(\lambda\hbar^3)$, has to occur before the time of equality, 
when $\rho\sim 5\times 10^{13}\,\textrm{eV cm}^{-3}$, we obtain $r_{\rm max}< 80\, \rm kpc$, 
i.e. $m/\lambda^{1/4}> 0.8\,\textrm{eV/c}^2$. However, the number of extra relativistic species at Big-Bang nucleosynthesis
places tighter constraints on the parameters of the theory. 
For a scalar field, this quantity, defined as the number of extra relativistic neutrino degrees of freedom at Big-Bang nucleosynthesis, takes the form
\begin{equation}
 \Delta N_{\rm eff} = 57.83 \times \left(\Omega_{\textrm{dm}}h^2\right)^{4/3}\left(\frac{\lambda^{1/4}}{m[\textrm{eV}/c^2]}\right)^{4/3}\,,
\end{equation}
see e.g. Eq.~(67) in Ref.~\cite{Arbey:2003sj}. (Note that there is an extra factor of 1/2 in our expression for $\Delta N_{\rm eff}$ with respect to that 
in Ref.~\cite{Arbey:2003sj}; this might come from the two helicities of the neutrino.)
Using the latest cosmological data provided by PLANCK+WP+highL~\cite{Ade:2013zuv},
$\Omega_{\textrm{dm}}h^2=0.1142\pm 0.0035$ at 1$\sigma$ C.L., and PLANCK+WP+highL+(D/H)$_{\rm p}$~\cite{Cooke:2013cba},
$\Delta N_{\rm eff}=0.23\pm 0.28$, also at 1$\sigma$ CL, we obtain
\begin{equation}
 r_{\textrm{max}}\lesssim 3\, \textrm{kpc}\,,\quad {\rm i.e.}\quad m/\lambda^{1/4}\gtrsim 4\,\textrm{eV}/c^2\,,
\end{equation}
Note that this result excludes marginally previous values of $r_{\rm max}\gtrsim 5.5\,\rm kpc$ 
arising from the study of the rotational curves of LSB and dwarf galaxies~\cite{Boehmer:2007um, Arbey:2003sj, Harko.2011, Robles.2012}.

The analysis in this paper applies only for the case of a self-interacting scalar particle with $\lambda> 0$; however,
similar results are expected when $\lambda\le 0$. Up to our knowledge, 
there is no analytic expression for the mass density profile of the halo model when the self-interaction term
is less than or equal to zero, but e.g. in the case of $\lambda=0$, 
the characteristic size and mass of the equilibrium configurations are found to be \cite{Ruffini} of order 
$R\sim Q^{-1/2}(m_{\textrm{Planck}}/m)(\hbar/mc)$,
and $M\sim Qm$, respectively. One can fix the number of particles, $Q$, and mass parameter, $m$, in order to describe the dynamics of dSphs,
implying $R\sim 1\, \textrm{kpc}$ and $M\sim 10^8\,M_{\odot}$,
see for instance Ref.~\cite{Lora} for the case of Ursa Minor,
but then configurations heavier than $10^8\,M_{\odot}$ would be smaller than $1\, \textrm{kpc}$, 
whereas those larger than $1\, \textrm{kpc}$ would result in halos lighter than $10^8\,M_{\odot}$.

In summary, if we dismiss previous constraints,
a scenario where the DM galactic halos are described by a single condensate is consistent with the data from the smallest and most DM-dominated nearby 
galactic systems; nonetheless, these single objects alone will not be consistent with the description of bigger galaxies.

\acknowledgments

We are grateful to Matthew Walker for helpful comments and discussions, and for providing us with the observational data.
We are also grateful to Juan Barranco and Octavio Valenzuela for useful comments on the first draft of this paper.
This work is supported by the UC MEXUS-CONACYT postdoctoral fellowship granted to AXGM. 
ADT is partly supported by CONACyT Mexico under Grants No. 182445 and No. 167335, and SP by
the U.S. Department of Energy under Contract No. DE-FG02-04ER41268.


\begin{thebibliography}{10}  

\bibitem{Springel:2005nw} V.~Springel {\it et al}, ``Simulating the joint evolution of quasars, galaxies and their large-scale distribution,''
                  Nature {\bf 435} 629-636 (2005) [astro-ph/0504097];
                  M.~Boylan-Kolchin, V.~Springel, S.D.M.~White, A.~Jenkins and G.~Lemson,
                  ``Resolving cosmic structure formation with the Millennium-II simulation,''
                  Mon.~Not.~Roy.~Astron.~Soc.  {\bf 398} 1150 (2009) [arXiv:0903.3041 [astro-ph.CO]];
                  A.~Klypin, S.~Trujillo-Gomez and J.~Primack,
                  ``Halos and galaxies in the standard cosmological model: results from the Bolshoi simulation,''
                  Astrophys. J. {\bf 740} 102 (2011) [arXiv:1002.3660 [astro-ph.CO];
                  K.~Riebe {\it et al}, ``The MultiDark database: Release of the Bolshoi and MultiDark cosmological simulations,''
                  (2011) [arXiv:1109.0003[astro-ph.CO]];
                  J.H~Kim {\it et al}, ``The AGORA high-resolution galaxy simulations comparison project,''
                  Astrophys.~J.~Suppl. {\bf 210} 14 (2014) [arXiv:1308.2669[astro-ph.GA]]
                  
\bibitem{direct} C.~Kelso, D.~Hooper and M.R.~Buckley, ``Toward a consistent picture for CRESST, CoGeNT and DAMA,''
                  Phys. Rev. D {\bf 85} 043515 (2012) [arXiv:1110.5338]; 
                 C.~Arina, ``Chasing a consistent picture for dark matter direct searches,''
                  Phys. Rev. D {\bf 86} 123527 (2012) [arXiv:1210.4011]

\bibitem{indirect} A.~Hektor, M.~Raidal and E.~Tempel, ``An evidence for indirect detection of dark matter from galaxy clusters in Fermi-LAT data,''
                   Astrophys. J. {\bf 762} L22 (2013) [arXiv:1207.4466];
                   S.~Profumo, ``Dissecting cosmic-ray electron-positron data with Occam's Razor: the role of known pulsars,''
                   Central Eur. J. Phys. {\bf 10} 1-31 (2011) [arXiv:0812.4457]
                   
\bibitem{WIMP}  G.~Bertone, D.~Hooper and J.~Silk, ``Particle dark matter: Evidence, candidates and constraints,''
                Phys. Rept. {\bf 405} 279-390 (2005) [hep-ph/0404175]
                
\bibitem{asymmetricDM} K.~Petraki and R.R.~Volkas, ``Review of asymmetric dark matter,''
                       Int. J. Mod. Phys. A {\bf 28} 1330028 (2013) [arXiv:1305.4939[hep-ph]]
                       
\bibitem{misalignment} M.~Dine and W.~Fischler, ``The not so harmless axion,''
                       Phys. Lett. B {\bf 120}) 137-141 (1983)

\bibitem{Suarez:2013iw} A.~Suarez, V.~Robles and T.~Matos, ``A review on the Scalar Field/Bose-Einstein Condensate Dark Matter Model,''
                   [arXiv:1302.0903[astro-ph.CO]]   
        
\bibitem{Boehmer:2007um} C.G.~Boehmer and T.~Harko, ``Can dark matter be a Bose-Einstein condensate?,''
        JCAP {\bf 0706} 025 (2007) [arXiv:0705.4158[astro-ph]]  
        
\bibitem{Qball} J.~Frieman, G.~Gelmini, M.~Gleiser and E.~Kolb, ``Solitogenesis: Primordial origin of nontopological solitons,''
                Phys. Rev. Lett. {\bf 60} 2101 (1988); 
                A.~Kusenko and M.~Shaposhnikov, ``Supersymmetric Q balls as dark matter,''
                Phys. Lett. B {\bf 418} 46–54 (1998) [hep-ph/9709492]
                   
\bibitem{Hu:2000ke} W.~Hu, R.~Barkana and A.~Gruzinov, ``Cold and fuzzy dark matter,''
        Phys. Rev. Lett. {\bf 85} 1158-1161 (2000) [astro-ph/0003365]                   
        
\bibitem{Sin-Bose} S.J.~Sin, ``Late-time phase transition and the galactic halo as a Bose liquid,''
        Phys. Rev. D {\bf 50} 3650 (1994);       
                   
\bibitem{Peebles:2000yy} P.J.E.~Peebles, ``Fluid dark matter,''
        [astro-ph/0002495]                                                 
                   
\bibitem{BECgeneral}  
        V.~Sahni and L.~Wang, ``A new cosmological model of quintessence and dark matter,''
        Phys. Rev. D {\bf 62} 103517 (2000) [astro-ph/9910097];
        J.~Goodman, ``Repulsive dark matter,'' (2000) [astro-ph/0003018];
        M.~Alcubierre, F.S.~Guzman, T.~Matos, D.~Nunez, L.A.~Urena-Lopez and P.~Wiederhold, ``Galactic collapse of scalar field dark matter,''
        Class. Quant. Grav. {\bf 19} 5017 (2002) [gr-qc/0110102];
        J.W.~Lee, ``Is dark matter a BEC or scalar field?,''
        J. Korean Phys. Soc. {\bf 54} 2622 (2009) [arXiv:0801.1442[astro-ph]];
        T.~Rindler-Daller and P.R.~Shapiro, ``Complex scalar field dark matter on galactic scales,''
        Mod. Phys. Lett. A {\bf 29} 1430002 (2014) [arXiv:1312.1734[astro-ph.CO]]
              
\bibitem{Sikivie:2009qn} P.~Sikivie, ``Axion cosmology,'' Lect. Notes Phys. {\bf 741} 19-50 (2008) [astro-ph/0610440] 
        P.~Sikivie and Q.~Yang, ``Bose-Einstein condensation of dark matter axions,''
        Phys. Rev. Lett. {\bf 103} 111301 (2009) [arXiv:0901.1106[hep-ph]] 
        
\bibitem{Arbey:2002} A.~Arbey, J.~Lesgourgues and P.~Salati, ``Cosmological constraints on quintessential halos,''
                     Phys. Rev. D {\bf 65} 083514 (2002) [astro-ph/0112324] 
        
\bibitem{Arbey:2003sj} A.~Arbey, J.~Lesgourgues and P.~Salati, ``Galactic halos of fluid dark matter,''    
        Phys. Rev. D {\bf 68} 023511 (2003) [astro-ph/0301533]
\bibitem{Lee:1995af} J.W.~Lee and I.G.~Koh, ``Galactic halos as boson stars,''
        Phys. Rev. D {\bf 53} 2236-2239 (1996) [hep-ph/9507385]        
\bibitem{Harko.2011} T.~Harko, ``Bose-Einstein condensation of dark matter solves the core/cusp problem,''
        JCAP {\bf 1105} 022 (2011) [arXiv:1105.2996[astro-ph.CO]]
\bibitem{Robles.2012} V.H.~Robles and T.~Matos, ``Flat central density profile and constant DM surface density in galaxies from Scalar Field Dark Matter,'' 
        Mon. Not. Roy. Astron. Soc. {\bf 422} 282-289 (2012) [arXiv:1201.3032[astro-ph.CO]]
\bibitem{Lora} V.~Lora, J.~Magana, A.~Bernal, F.J.~Sanchez-Salcedo and E.K.~Grebel, ``On the mass of ultra-light bosonic dark matter from galactic dynamics,'' 
        JCAP {\bf 1202} 011 (2012) [arXiv:1110.2684[astro-ph.GA]] 
\bibitem{Lora2}  V.~Lora and J.~Magana, ``Is Sextans dwarf galaxy in a scalar field dark matter halo?,'' (2014) [arXiv:1406.6875]              
\bibitem{Dwornik:2013fra} M.~Dwornik, Z.~Keresztes and L.A.~Gergely, ``Rotation curves in Bose-Einstein condensate dark matter halos,'' 
        (2013) [arXiv:1312.3715[gr-qc]]  
\bibitem{Hua} M.H.~Li and Z.B.~Li, ``Constraints on Bose-Einstein-condensed axion dark matter from the Hi nearby galaxy survey data,''
        Phys. Rev. D {\bf 89} 103512 (2014) [arXiv:1406.1312[astro-ph.GA]]
        
\bibitem{Peccei:1977hh} R.D.~Peccei and H.R.~Quinn, ``CP conservation in the presence of pseudoparticles,''
        Phys. Rev. Lett. {\bf 38} 1440 (1977)
        
\bibitem{Barranco:2010ib} M.C.~Johnson and M.~Kamionkowski, ``Dynamical and gravitational instability of oscillating-field dark energy and dark matter,''
        Phys. Rev. D {\bf 78} 063010 (2008) [arXiv:0805.1748[astro-ph]];
        J.~Barranco and A.~Bernal, ``Self-gravitating system made of axions,''
        Phys. Rev. D {\bf 83} 043525 (2011) [arXiv:1001.1769[astro-ph.CO]]    
        
\bibitem{GonzalezMorales:2012ab} A.X.~Gonzalez-Morales, O.~Valenzuela and L.A.~Aguilar, ``Constraining dark matter sub-structure with the dynamics of astrophysical systems,''
        JCAP {\bf 1303} 001 (2013) [arXiv:1211.6745[astro-ph.CO]];
        F.~Iocco, M.~Pato, G.~Bertone and P.~Jetzer, ``Dark matter distribution in the Milky Way: Microlensing and dynamical constraints,''
        JCAP {\bf 1111} 029 (2011) [arXiv:1107.5810[astro-ph.GA]]

\bibitem{Colpi:1986ye} M.~Colpi, S.L.~Shapiro and I.~Wasserman, ``Boson stars: gravitational equilibria of selfinteracting scalar fields,''
        Phys. Rev. Lett. {\bf 57} 2485-2488 (1986)
        
\bibitem{Bilic:2000ef} N.~Bilic and H.~Nikolic, ``Selfgravitating bosons at nonzero temperature,''
        Nucl. Phys. B {\bf 590} 575-595 (2000) [gr-qc/0006065]        
        
\bibitem{Mateo:1998wg} M.~Mateo, ``Dwarf galaxies of the Local Group,''
        Ann. Rev. Astron. Astrophys., {\bf 36} 435-506 (1998) [astro-ph/9810070]         
        
\bibitem{Walker:2009zp} M.G.~Walker, M.~Mateo, E.W.~Olszewski, J.~Pe\~narrubia, N.W.~Evans and G.~Gilmore, ``A universal mass profile for dwarf spheroidal galaxies,''
        Astrophys. J. {\bf 704} 1274-1287 (2009) [arXiv:0906.0341[astro-ph.CO]];  Erratum-ibid. {\bf 710} 886-890 (2010)    
        
\bibitem{Salucci:2011ee} P.~Salucci, M.I.~Wilkinson, M.G.~Walker, G.F.~Gilmore, E.K.~Grebel, A.~Koch, C.F.~Martins and R.F.G.~Wyse,
        ``Dwarf spheroidal galaxy kinematics and spiral galaxy scaling laws,''
        Mon. Not. Roy. Astron. Soc., {\bf 420} 2034-2041 (2012) [arXiv:1111.1165[astro-ph.CO]]  
        
\bibitem{jardel} J.R.~Jardel and K.~Gebhardt, ``Variations in a universal dark matter profile for dwarf spheroidals,''
                 Astrophys. J. {\bf 775}  L30 (2013) 
                 
\bibitem{Valenzuela:2005dh}  O.~Valenzuela, G.~Rhee, A.~Klypin, F.~Governato, G.~Stinson, T.~R.~Quinn and J.~Wadsley, 
                             ``Is there evidence for flat cores in the halos of dwarf galaxies?: the case of ngc 3109 and ngc 6822,''   
                             Astrophys. J. {\bf 657}, 773 (2007) [astro-ph/0509644].

\bibitem{Walker:2007ju} M.G.~Walker, M.~Mateo, E.W.~Olszewski, O.Y.~Gnedin, X.~Wang, B.~Sen and M.~Woodroofe, ``Velocity dispersion profiles of seven dwarf spheroidal galaxies,''
        Astrophys. J. {\bf 667} L53-L56 (2007) [arXiv:0708.0010[astro-ph]]    
               
\bibitem{Mateo2008} M.~Mateo, E.W.~Olszewski and M.G.~Walker, ``The velocity dispersion profile of the remote dwarf spheroidal galaxy Leo I: A tidal hit and run?,''
        Astrophys. J. {\bf 675} 201-233 (2008) [arXiv:0708.1327[astro-ph]];
        M.G.~Walker, M.~Mateo and E.W.~Olszewski, ``Stellar velocities in the Carina, Fornax, Sculptor, and Sextans dSph galaxies: Data from the Magellan/MMFS Survey,''
        Astrophys. J. {\bf 137} 3100-3108 (2009) [arXiv:0708.1327[astro-ph]]
        
\bibitem{Binney} J.~Binney and S.~Tremaine, Galactic Dynamics, Princeton University Press, 2 edition (January 27, 2008) 904pp 

\bibitem{Sersic} J.L.~Sersic, ``Atlas de galaxias australes,'' Cordoba, Argentina: Observatorio Astronomico (1968)

\bibitem{King} I.~King, ``The structure of star clusters. I. an empirical density law,''
               Astron. J. {\bf 67} 471 (1962)
               
\bibitem{Foreman-Mackey} D.~Foreman-Mackey, D.W.~Hogg, D.~Lang and J.~Goodman, ``emcee: The MCMC Hammer,'' 
                         Publications of the Astronomical Society of the Pacific {\bf 125} 306-312 (2013) [arXiv:1202.3665[astro-ph.IM]]

              
\bibitem{Jardel:2012am} J.R.~Jardel, K.~Gebhardt, M.H.~Fabricius, N.~Drory and M.J.~Williams, 
                       ``Measuring dark matter profiles non-parametrically in dwarf spheroidals: An application to Draco,''
                       Astrophys. J. {\bf 763} 91 (2013) [arXiv:1211.5376[astro-ph.CO]] 
                       
\bibitem{Gerhard} O.~Gerhard and D.N.~Spergel, ``Dwarf spheroidal galaxies and the mass of the neutrino,''
                  Astrophys. J. {\bf 389} L9-L11 (1992)                  
        
\bibitem{Lokas2008} E.L.~Lokas, J.~Klimentowski, S.~Kazantzidis and L.~Mayer, ``The anatomy of Leo I: How tidal tails affect the kinematics,''
        Mon. Not. Roy. Astron. Soc., {\bf 390} 625-634 (2008) [arXiv:0804.0204[astro-ph]];
        E.L.~Lokas, ``The mass and velocity anisotropy of the Carina, Fornax, Sculptor and Sextans dwarf spheroidal galaxies,''
        Mon. Not. Roy. Astron. Soc., {\bf 394} L102-L106 (2009) [arXiv:0901.0715[astro-ph.GA]]
        
\bibitem{Ade:2013zuv}  P.A.R.~Ade {\it et al.} [Planck Collaboration], ``Planck 2013 results. XVI. Cosmological parameters,'' [arXiv:1303.5076[astro-ph.CO]]  

\bibitem{Cooke:2013cba} R.~Cooke, M.~Pettini, R.~A.~Jorgenson, M.~T.~Murphy and C.~C.~Steidel, ``Precision measures of the primordial abundance of deuterium,''
  [arXiv:1308.3240[astro-ph.CO]]
        
\bibitem{Ruffini} R.~Ruffini and S.~Bonazzola, ``Systems of self-gravitating particles in general relativity and the concept of an equation of state,''
        Phys. Rev. {\bf 187} 1767-1783 (1969)         
        
        
        
\end{thebibliography}
\end{document}